\documentclass[preprintnumbers,prd,twocolumn,showpacs,floatfix,preprintnumbers,letterpaper,amsmath,amssymb,nofootinbib,superscriptaddress]{revtex4}
\usepackage{graphicx}
\usepackage{epsfig}
\usepackage{bm}
\usepackage{amsfonts}

\newcommand{\la}{\lambda}

\def\e{\epsilon}

\def\beq{\begin{equation}}
\def\eeq{\end{equation}}
\def\bea{\begin{eqnarray}}
\def\eea{\end{eqnarray}}
\def\bean{\begin{eqnarray*}}
\def\eean{\end{eqnarray*}}

\newcommand{\bear}{\begin{eqnarray}}

\newcommand{\eear}{\end{eqnarray}}

\newbox\pippobox

\def\6{\partial}

\def\a{\alpha}

\def\e{\epsilon}

\def\sq
\def\a{\alpha}

\def\l{\lambda}

\def\na{\nabla}

\def\e{\epsilon}

\def\dn{\Delta N_{\nu}}

\newcommand {\lla} {\ {\raise-.5ex\hbox{$\buildrel<\over\sim$}}\ }

\def\be{\begin{equation}}
\def\ee{\end{equation}}
\def\ba{\begin{eqnarray}}
\def\ea{\end{eqnarray}}
\def\w{\omega}
\renewcommand{\(}{\left(}
\renewcommand{\)}{\right)}
\renewcommand{\[}{\left[}
\renewcommand{\]}{\right]}

\begin{document}

\title{Aspects of Ho\v{r}ava-Lifshitz cosmology}

\author{Emmanuel N. Saridakis}
 \email{msaridak@phys.uoa.gr}
 \affiliation{College of Mathematics and Physics,\\ Chongqing University of Posts and
Telecommunications, Chongqing, 400065, P.R. China }

\begin{abstract}
We review some general aspects of Ho\v{r}ava-Lifshitz cosmology.
Formulating it in its basic version, we extract the cosmological
equations and we use observational data in order to constrain the
parameters of the theory. Through a phase-space analysis we
extract  the late-time stable solutions, and we show that eternal
expansion, and bouncing and cyclic behavior can arise naturally.
Concerning the effective dark energy sector we show  that it can
describe the phantom phase without the use of a phantom field.
However, performing a detailed perturbation analysis, we see that
Ho\v{r}ava-Lifshitz gravity in its basic version suffers from
instabilities. Therefore, suitable generalizations are required in
order for this novel theory to be a candidate for the description
of nature.
\end{abstract}

 \pacs{98.80.-k, 04.60.Bc, 04.50.Kd}

\maketitle

\section{Introduction}

Almost one year ago  Ho\v{r}ava proposed a power-counting
renormalizable theory with consistent ultra-violet (UV) behavior
\cite{hor2,hor3,hor1}. Although presenting an infrared (IR) fixed
point, namely General Relativity, in the  UV the theory exhibits
an anisotropic, Lifshitz scaling between time and space. Due to
these novel features, there has been a large amount of effort in
examining and extending the properties of the theory itself
\cite{Cai:2009ar,Charmousis:2009tc,Li:2009bg,
Sotiriou:2009bx,Bogdanos:2009uj,Kluson:2009rk,
Blas:2009qj,Kiritsis:2009vz}. Additionally, application of
Ho\v{r}ava-Lifshitz gravity as a cosmological framework gives rise
to Ho\v{r}ava-Lifshitz cosmology, which proves to lead to
interesting behavior \cite{Kiritsis:2009sh}. In particular, one
can examine specific solution subclasses
\cite{Lu:2009em,Minamitsuji:2009ii,Wu:2009ah}, the phase-space
behavior \cite{Carloni:2009jc,Leon:2009rc,Myung:2009if}, the
gravitational wave production \cite{Mukohyama:2009zs}, the
perturbation spectrum
\cite{Mukohyama:2009gg,Chen:2009jr,Cai:2009hc}, the matter bounce
\cite{Brandenberger:2009yt,Brandenberger:2009ic,Cai:2009in,Gao:2009wn},
the black hole properties
\cite{Danielsson:2009gi,Park:2009zra,Lee:2009rm}, the dark energy
phenomenology
\cite{Saridakis:2009bv,Park:2009zr,Chaichian:2010yi,Jamil:2010vr},
the observational constraints on the parameters of the theory
\cite{Dutta:2009jn,Dutta:2010jh,Chiang:2010js}, the astrophysical
phenomenology \cite{Kim:2009dq}, the thermodynamic properties
\cite{Wang:2009rw,Cai:2009qs} etc. However, despite this extended
research, there are still many ambiguities if Ho\v{r}ava-Lifshitz
gravity is reliable and capable of a successful description of the
gravitational background of our world, as well as of the
cosmological behavior of the universe
\cite{Charmousis:2009tc,Li:2009bg,Sotiriou:2009bx,Bogdanos:2009uj,Koyama:2009hc,Papazoglou:2009fj}.

In the present work we review the basic aspects of
Ho\v{r}ava-Lifshitz cosmology. The manuscript is organized as
follows: In section \ref{model} we present the simple version of
Ho\v{r}ava-Lifshitz cosmology, in both its detailed-balance and
beyond-detailed-balance version. In section \ref{Observational
constraints} we use observational data in order to constrain the
parameters of the theory. In section \ref{leonmanos} we present
the results of the phase-space analysis, in \ref{bouncecyclic} we
present the bouncing and  cyclic solutions, and in
 \ref{realde} we extend the theory in order to present a more
 realistic dark energy phenomenology. In section  \ref{instabil}, through a
 perturbation analysis, we discuss the instabilities of the simple
 versions of the theory, and thus in section
  \ref{healthy} we present a healthy extension of Ho\v{r}ava-Lifshitz gravity. Finally, section  \ref{conclusions}
 is devoted to the summary
of our results.

\section{Ho\v{r}ava-Lifshitz cosmology}
\label{model}

In this section we briefly review the scenario where the
cosmological evolution is governed by the simple version of
Ho\v{r}ava-Lifshitz gravity \cite{Kiritsis:2009sh}. The dynamical
variables are the lapse and shift functions, $N$ and $N_i$
respectively, and the spatial metric $g_{ij}$ (roman letters
indicate spatial indices). In terms of these fields the full
metric is written as:
\begin{eqnarray}\label{metriciit}
ds^2 = - N^2 dt^2 + g_{ij} (dx^i + N^i dt ) ( dx^j + N^j dt ) ,
\end{eqnarray}
and the scaling transformation of the coordinates reads as $
 t \rightarrow l^3 t~~~{\rm and}\ \ x^i \rightarrow l x^i
$.

\subsection{Detailed Balance}

The gravitational action is decomposed into a kinetic and a
potential part as $S_g = \int dt d^3x \sqrt{g} N ({\cal L}_K+{\cal
L}_V)$. The assumption of detailed balance \cite{hor3}
  reduces the possible terms in the Lagrangian, and it allows
for a quantum inheritance principle \cite{hor2}, since the
$(D+1)$-dimensional theory acquires the renormalization properties
of the $D$-dimensional one. Under the detailed balance condition
 the full action of Ho\v{r}ava-Lifshitz gravity is given by
\begin{eqnarray}
 S_g &=&  \int dt d^3x \sqrt{g} N \left\{
\frac{2}{\kappa^2}
(K_{ij}K^{ij} - \lambda K^2) \ \ \ \ \ \ \ \ \ \ \ \ \ \ \ \ \  \right. \nonumber \\
&+&\left.\frac{\kappa^2}{2 w^4} C_{ij}C^{ij}
 -\frac{\kappa^2 \mu}{2 w^2}
\frac{\epsilon^{ijk}}{\sqrt{g}} R_{il} \nabla_j R^l_k +
\frac{\kappa^2 \mu^2}{8} R_{ij} R^{ij}
     \right. \nonumber \\
&-&\left.    \frac{\kappa^2 \mu^2}{8( 3 \lambda-1)} \left[ \frac{1
- 4 \lambda}{4} R^2 + \Lambda  R - 3 \Lambda ^2 \right] \right\},
\label{acct}
\end{eqnarray}
where $ K_{ij} =   \left( {\dot{g_{ij}}} - \nabla_i N_j - \nabla_j
N_i \right)/2N $
 is the extrinsic curvature and
$ C^{ij}  =  \epsilon^{ijk} \nabla_k \bigl( R^j_i -  R
\delta^j_i/4 \bigr)/\sqrt{g} $ the Cotton tensor, and the
covariant derivatives are defined with respect to the spatial
metric $g_{ij}$. $\epsilon^{ijk}$ is the totally antisymmetric
unit tensor, $\lambda$ is a dimensionless constant and the
variables $\kappa$, $w$ and $\mu$ are constants. Finally, we
mention that in action (\ref{acct}) we have already performed the
usual analytic continuation of the parameters $\mu$ and $w$ of the
original version of Ho\v{r}ava-Lifshitz gravity, since such a
procedure is required in order to obtain a realistic cosmology
\cite{Lu:2009em,Minamitsuji:2009ii,Wang:2009rw,Park:2009zra}
(although it could fatally affect the gravitational theory
itself).

In order to add the matter component we follow the hydrodynamical
approach   of adding a cosmological stress-energy tensor to the
gravitational field equations, by demanding to recover the usual
general relativity formulation in the low-energy limit
\cite{Sotiriou:2009bx,Chaichian:2010yi,Carloni:2009jc}. Thus, this
matter-tensor is a hydrodynamical approximation with $\rho_m$ and
$p_m$ (or $\rho_m$ and $w_m$) as parameters. Similarly, one can
additionally include the standard-model-radiation component, with
the additional parameters   $\rho_r$ and $w_r$.

In order to investigate cosmological frameworks, we impose the
projectability condition \cite{Charmousis:2009tc} and we use an
FRW metric
\begin{eqnarray}
N=1~,~~g_{ij}=a^2(t)\gamma_{ij}~,~~N^i=0~,
\end{eqnarray}
with
\begin{eqnarray}
\gamma_{ij}dx^idx^j=\frac{dr^2}{1- K r^2}+r^2d\Omega_2^2~,
\end{eqnarray}
where $ K<,=,> 0$ corresponding  to open, flat, and closed
universe respectively. By varying $N$ and $g_{ij}$, we extract the
Friedmann equations:
\begin{eqnarray}\label{Fr1fluid}
H^2 &=&
\frac{\kappa^2}{6(3\la-1)}\Big(\rho_m+\rho_r\Big)+\nonumber\\
&+&\frac{\kappa^2}{6(3\la-1)}\left[ \frac{3\kappa^2\mu^2
K^2}{8(3\lambda-1)a^4} +\frac{3\kappa^2\mu^2\Lambda
^2}{8(3\lambda-1)}
 \right]-\nonumber\\
 &-&\frac{\kappa^4\mu^2\Lambda  K}{8(3\lambda-1)^2a^2} \ ,
\end{eqnarray}
\begin{eqnarray}\label{Fr2fluid}
\dot{H}+\frac{3}{2}H^2 &=&
-\frac{\kappa^2}{4(3\la-1)}\Big(w_m\rho_m+w_r\rho_r\Big)-\nonumber\\
&-&\frac{\kappa^2}{4(3\la-1)}\left[\frac{\kappa^2\mu^2
K^2}{8(3\lambda-1)a^4} -\frac{3\kappa^2\mu^2\Lambda
^2}{8(3\lambda-1)}
 \right]-\nonumber\\
 &-&\frac{\kappa^4\mu^2\Lambda  K}{16(3\lambda-1)^2a^2}\ ,
\end{eqnarray}
where  $H\equiv\frac{\dot a}{a}$ is the Hubble parameter. As
usual, $\rho_m$ follows the standard evolution equation
$
 \dot{\rho}_m+3H(\rho_m+p_m)=0,
$ while $\rho_r$   follows $
 \dot{\rho}_r+3H(\rho_r+p_r)=0.
$
Finally, concerning the dark-energy sector we can define
\begin{equation}\label{rhoDE}
\rho_{DE}\equiv \frac{3\kappa^2\mu^2 K^2}{8(3\lambda-1)a^4}
+\frac{3\kappa^2\mu^2\Lambda ^2}{8(3\lambda-1)}
\end{equation}
\begin{equation}
\label{pDE} p_{DE}\equiv \frac{\kappa^2\mu^2
K^2}{8(3\lambda-1)a^4} -\frac{3\kappa^2\mu^2\Lambda
^2}{8(3\lambda-1)}.
\end{equation}
The term proportional to $a^{-4}$ is the usual ``dark radiation
term'', present in Ho\v{r}ava-Lifshitz cosmology
\cite{Kiritsis:2009sh}, while the constant term is just the
explicit cosmological constant. Therefore, in expressions
(\ref{rhoDE}),(\ref{pDE}) we have defined the energy density and
pressure for the effective dark energy, which incorporates the
aforementioned contributions. Note that using
(\ref{rhoDE}),(\ref{pDE}) it is straightforward to show that these
 dark energy quantities satisfy the
standard evolution equation:
$\dot{\rho}_{DE}+3H(\rho_{DE}+p_{DE})=0.
$

If we require expressions (\ref{Fr1fluid}) to coincide with the
standard Friedmann equations, in units where $c=1$  we set
\cite{Kiritsis:2009sh}:
\begin{eqnarray}
G_{\rm cosmo}&=&\frac{\kappa^2}{16\pi(3\lambda-1)}\nonumber\\
\frac{\kappa^4\mu^2\Lambda}{8(3\lambda-1)^2}&=&1,
\label{simpleconstants0}
\end{eqnarray}
where $G_{\rm cosmo}$ is the ``cosmological'' Newton's constant,
that is the one that is read from the Friedmann equations. We
mention that in theories with Lorentz invariance breaking $G_{\rm
cosmo}$ does not coincide with the
 ``gravitational'' Newton's constant
$G_{\rm grav}$, that is the one that is read from the action,
unless Lorentz invariance is restored \cite{Carroll:2004ai}. For
completeness we mention that in our case $ G_{\rm
grav}=\kappa^2/(32\pi)$, as it can be straightforwardly read from
the action (\ref{acct}). Thus, it becomes obvious that in the IR
($\lambda=1$), where Lorentz invariance is restored, $G_{\rm
cosmo}$ and $G_{\rm grav}$ coincide.

\subsection{Beyond Detailed Balance}

The aforementioned formulation of Ho\v{r}ava-Lifshitz cosmology
has been performed under the imposition of the detailed-balance
condition. However, in the literature there is a discussion
whether this condition leads to reliable results or if it is able
to reveal the full information of Ho\v{r}ava-Lifshitz
 gravity \cite{Kiritsis:2009sh}. Therefore, one
 needs to investigate also the Friedman equations in the case
 where detailed balance is relaxed. In such a case one can in
 general write
 \cite{Charmousis:2009tc,Sotiriou:2009bx,Bogdanos:2009uj,Carloni:2009jc,Leon:2009rc}:
\begin{eqnarray}\label{Fr1c}
H^2 &=&
\frac{2\sigma_0}{(3\la-1)}\Big(\rho_m+\rho_r\Big)+\nonumber\\
&+&\frac{2}{(3\la-1)}\left[ \frac{\sigma_1}{6}+\frac{\sigma_3
K^2}{6a^4} +\frac{\sigma_4 K}{6a^6}
 \right]+\nonumber\\&+&\frac{\sigma_2}{3(3\la-1)}\frac{ K}{a^2}
\end{eqnarray}
\begin{eqnarray}\label{Fr2c}
\dot{H}+\frac{3}{2}H^2 &=&
-\frac{3\sigma_0}{(3\la-1)}\Big(w_m\rho_m+w_r\rho_r\Big)-\nonumber\\
&-&\frac{3}{(3\la-1)}\left[ -\frac{\sigma_1}{6}+\frac{\sigma_3
K^2}{18a^4} +\frac{\sigma_4 K}{6a^6}
 \right]+\nonumber\\&+&
 \frac{\sigma_2}{6(3\la-1)}\frac{ K}{a^2},
\end{eqnarray}
where $\sigma_0\equiv \kappa^2/12$, and the constants $\sigma_i$
are arbitrary (with $\sigma_2$ being negative and $\sigma_4$
positive). Furthermore, the   dark-energy quantities
 are generalized to
\begin{eqnarray}\label{rhoDEext}
&&\rho_{DE}|_{_\text{non-db}}\equiv
\frac{\sigma_1}{6}+\frac{\sigma_3 K^2}{6a^4} +\frac{\sigma_4
K}{6a^6}
\\
&&\label{pDEext} p_{DE}|_{_\text{non-db}}\equiv
-\frac{\sigma_1}{6}+\frac{\sigma_3 K^2}{18a^4} +\frac{\sigma_4
K}{6a^6}.
\end{eqnarray}
Again, it is easy to show that
\begin{eqnarray}\label{rhodotfluidnd}
\dot{\rho}_{DE}|_{_\text{non-db}}+3H(\rho_{DE}|_{_\text{non-db}}+p_{DE}|_{_\text{non-db}})=0.
\end{eqnarray}
Finally, if we force (\ref{Fr1c}),(\ref{Fr2c}) to coincide with
 the standard Friedmann equations, we obtain:
\begin{eqnarray}
&&G_{\rm cosmo}=\frac{6\sigma_0}{8\pi(3\lambda-1)}\nonumber\\
&&\sigma_2=-3(3\lambda-1), \label{simpleconstants0nd}
\end{eqnarray}
while in this case the ``gravitational'' Newton's constant $G_{\rm
grav}$ writes as $ G_{\rm grav}= 6\sigma_0/(16\pi)$. Similarly to
the detailed balance case, in the IR ($\lambda=1$) $G_{\rm cosmo}$
and $G_{\rm grav}$ coincide.

\section{Observational constraints}
\label{Observational constraints}

Having presented the cosmological equations of a universe governed
by Ho\v{r}ava-Lifshitz gravity, both with and without the
detailed-balance condition, we now proceed to study  the
observational constraints on the model parameters
\cite{Dutta:2009jn,Dutta:2010jh}.

\subsection{Constraints on Detailed-Balance scenario}

We work in the usual units suitable for observational comparisons,
namely setting  $8\pi G_{\rm grav}=1$ (we have already set $c=1$
in order to obtain (\ref{simpleconstants0})). This allows us to
reduce the parameter space, since in this case $ G_{\rm grav}$
gives $ \kappa^2=4$ and thus (\ref{simpleconstants0}) lead to: $
G_{\rm cosmo}=\frac{1}{4\pi(3\lambda-1)}$ an
$\mu^2\Lambda=\frac{(3\lambda-1)^2}{2}$. In order to proceed to
the elaboration of observational data,  we consider as usual the
matter (dark plus baryonic) component to be dust, that is
$w_m\approx0$, and similarly for the standard-model radiation we
consider $w_r=1/3$, where both assumptions are valid in the epochs
in which observations focus. Therefore, the corresponding
evolution equations
  give
$\rho_m=\rho_{m0}/a^3$ and $\rho_r=\rho_{r0}/a^4$ respectively.
Additionally, instead of the scale factor it proves convenient to
use the redshift $z$ as the independent variable, which is given
by $1+z\equiv a_0/a=1/a$. Finally, we introduce the usual density
parameters ($\Omega_m\equiv\rho_m/(3H^2)$, $\Omega_ K\equiv -
K/(H^2a^2)$, $\Omega_r\equiv\rho_r/(3H^2)$).
 Inserting these relations into Friedmann equation
(\ref{Fr1fluid}) we acquire:
\begin{eqnarray}
H^2&=&H_{0}^2\Big\{\frac{2}{(3\lambda-1)}\Big[\Omega_{m0}(1+z)^3+\Omega_{r0}(1+z)^4\Big]+\nonumber\\
&\ &+\Omega_{ K0}(1+z)^2+\Big[\omega+\frac{\Omega_{ K
0}^2}{4\omega}(1+z)^4\Big] \Big\},
 \label{Frdbfinal}
\end{eqnarray}
  where we have also introduced  the
dimensionless parameter $ \omega\equiv\frac{\Lambda}{2 H_0^2}$,
and where
 a $0$-subscript denotes the present value of the
corresponding quantity. Applying this relation at present we get:
  \be
\label{cond1}
\frac{2}{(3\lambda-1)}\Big(\Omega_{m0}+\Omega_{r0}\Big)+\Omega_{K0}+\omega+\frac{\Omega_{K0}^2}{4\omega}=1.
 \ee

We remind that the term $\Omega_{ K 0}^2/(4\omega)$ is the
coefficient of the dark radiation term, which is a characteristic
feature of the Ho\v{r}ava-Lifshitz gravitational background. Since
this dark radiation component has been present also during the
time of nucleosynthesis, it is subject to bounds from Big Bang
Nucleosynthesis (BBN). As discussed in more details in the
Appendix of \cite{Dutta:2009jn}, if the upper limit on the total
amount of dark radiation allowed during BBN is expressed through
the parameter $\Delta N_\nu$ of the effective neutrino species
\cite{BBNrefs}, then we obtain the following constraint:
  \be
\label{cond2}
 \frac{\Omega_{ K 0}^2}{4\omega}=0.135\dn \Omega_{r0}.
  \ee

In summary, the scenario at hand involves four parameters (we fix
$H_0$ by its 7-year WMAP best-fit values, given in Table 1 of
\cite{Komatsu:2010fb}), namely $\Omega_{m0}$, $\Omega_{K0}$,
$\omega$ and $\dn$, subject to constraint  equations (\ref{cond1})
and (\ref{cond2}). We marginalize over the cosmological parameters
$\Omega_{m0}$, $\Omega_{b0}$, $\Omega_{r0}$ and $H_0$. Of the four
remaining parameters, only two are independent, and we choose
$\lambda$ and $\dn$  as our free parameters. Once these are
chosen, and for a given choice of curvature, $\Omega_{K0}$ and
$\omega$ are immediately fixed from the constraint equations. In
particular, $\omega$ can be determined by eliminating
$\Omega_{K0}$ from relations (\ref{cond1}) and (\ref{cond2}):
\begin{align}
\label{omega}
\omega-&2\,{\rm sgn}\(\Omega_{K0}\)\sqrt{0.135\dn\,\Omega_{r0}\,\omega}\,+\,0.135\dn\Omega_{r0}\nonumber\\
+\,&2\[\frac{\Omega_{m0}+\Omega_{r0}}{3\lambda-1}\]-1=0.
\end{align}
$\Omega_{K0}$ can then be found from $\omega$ using (\ref{cond2}).

\begin{figure}[ht]
\begin{center}
\includegraphics[width=6cm]{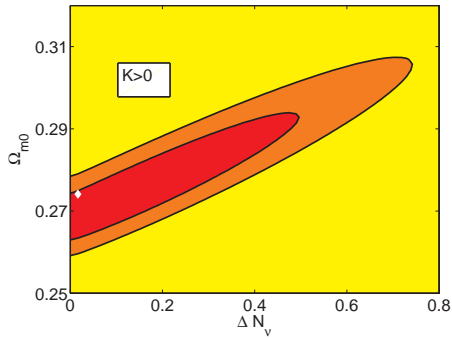}
\caption{(Color Online) {\it{ 1$\sigma$ and  2$\sigma$ contour
plots of $\Omega_{m0}$ vs
        $\dn$ for   positive curvature ($K>0$), under SNIa, BAO and CMB observational data.
             The white diamond marks the
best-fit point. The model parameters $\omega\equiv\Lambda/(2
H_0^2)$ and $\Omega_{K0}\equiv - K/(H^2_0)$ are related to
$\Omega_{m0}$ and $\dn$ through  (\ref{cond2}) and (\ref{omega}).
}}} \label{posdb}
\end{center}
\end{figure}

\begin{figure}[htbp]
\begin{center}
\includegraphics[width=6cm]{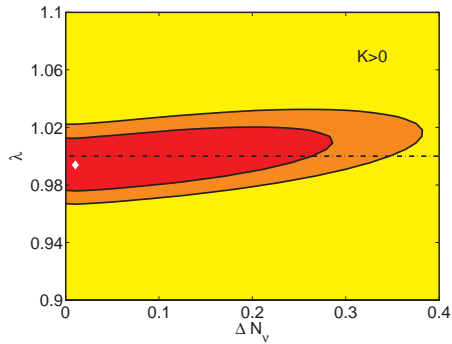}
\caption{ (Color online) {\it{1$\sigma$ and  2$\sigma$ contour
plots of $\lambda$ vs
        $\dn$ for positive
        ($K>0$) curvature, in the detailed-balance scenario,
        under SNIa, BAO and CMB observational data. The remaining
         parameters have been marginalized over.
          }}}
 \label{dbcontour}
 \end{center}
\end{figure}

In Fig. \ref{posdb}   we use a combination of observational data
from SNIa, BAO and CMB to construct  likelihood contours for the
parameters $\Omega_{m0}$ and $\dn$ for positive curvature.
Additionally, in Fig.~\ref{dbcontour} we display the
likelihood-contours for the free parameters $\lambda$ vs $\dn$ for
  positive   curvature, where all other parameters
have been marginalized over.
  Finally,  in
Table \ref{dblimits} we summarize the $1\sigma$ limits on the
parameter values for the detailed-balance scenario.
\begin{center}
\begin{table}[ht]
    \centering
    \scalebox{.95}{
        \begin{tabular}{|c|c|c|c|c|c|c|}
        \hline \textbf{$K$} &
         $\kappa^2/(8\pi G_{\rm grav})$ & \textbf{$\left(1/H_0^2\right)\Lambda $}& $ \left(8\pi G_{\rm grav} H_0\right)\mu$ & $\lambda$ & $\dn$\\\hline
        $>$0&     4      & $(0,\,1.46 )$ & $(1.37,\infty)$ & (0.98, 1.01) & (0, 0.32)\\\hline
           $<$0 &   4      & $(0,\,1.46) $ & $(1.8,\infty)$ & (0.97, 1.01)& (0, 0.68) \\\hline
        \end{tabular}}
    \caption{1$\sigma$ limits on the parameter values for the detailed-balance scenario, for positive and negative curvature.The cosmological parameters $\Omega_{m0}$, $\Omega_{b0}$, $\Omega_{r0}$ and $H_0$ have been marginalized over.}
    \label{dblimits}
\end{table}
\end{center}

In conclusion, we see that the Ho\v{r}ava-Lifshitz cosmological
scenario under the detailed balance condition is not ruled out by
observations. However, there are tight constraints on the model
parameters. Furthermore, the data constrain $\lambda$ to roughly
$\lambda=1^{+0.01}_{-0.02}$ at the $1\sigma$ level, that is to a
very narrow window around its IR value, while its best fit value
is very close to $1$ ($\lambda_{b.f}=0.006$).

\subsection{Constraints on Beyond-Detailed-Balance scenario}

In units where $8\pi G_{\rm grav}=1$, $ G_{\rm grav}$ gives $
\sigma_0=1/3$. Following the procedure of the previous subsection,
the Friedmann equation (\ref{Fr1c}) can be written as{\small{
 \ba
   \label{Fr1c_a}
&&H^2=H_{0}^2\Big\{\frac{2}{(3\lambda-1)}\Big[\Omega_{m0}(1+z)^3+\Omega_{r0}(1+z)^4\Big]+\nonumber\\
 &&\Omega_{ K0}\(1+z\)^2
+\frac{2}{(3\lambda-1)}\Big[\omega_1+\omega_3 \(1+z\)^4+\omega_4
\(1+z\)^6\Big]\Big\},\nonumber
  \ea}}
where we have introduced the dimensionless
 parameters $
\omega_1=\frac{\sigma_1}{6H_0^2}$, $ \omega_3=\frac{\sigma_3
H_0^2\Omega_{ K0}^2}{6}$ and $ \omega_4=-\frac{\sigma_4\Omega_{
K0}}{6}$.
   Additionally, we consider the
combination $\omega_4$ to be positive, in order to ensure that the
Hubble parameter is real for all redshifts.

In summary, the present scenario involves the following
parameters: the cosmological parameters $H_0$, $\Omega_{m0}$,
 $\Omega_{K0}$, $\Omega_{b0}$, $\Omega_{r0}$, and the model parameters $\lambda$, $\w_1$, $\w_3$
and $\w_4$. Similarly to the detailed-balance section these are
subject to two constraints. The first one arises from the Friedman
equation at $z=0$, which leads to
  \be
     \label{ndbcond1}
\frac{2}{(3\lambda-1)}\Big[\Omega_{m0}+\Omega_{r0}+\w_1+\w_3+\w_4\Big]+\Omega_{K0}=1.
  \ee
  This constraint eliminates the parameter $w_1$.
The second one arises from BBN considerations, since, as we show
in the Appendix of \cite{Dutta:2009jn}, at the time of BBN
($z=z_{\rm BBN}$) we acquire \cite{BBNrefs}:
 \be
\label{ndbcond2} \w_3+\w_4\(1+z_{\rm
BBN}\)^2=\w_{3\text{max}}\equiv0.135\dn\Omega_{r0}, \ee   where
$\w_{3\text{max}}$ denotes the upper limit on $\w_3$. In the
following, we use expression (\ref{ndbcond2}) to eliminate $\w_4$.
For convenience, instead of $\w_3$ we define the new parameter $
\alpha\equiv  \frac{\w_3}{\w_{3\text{max}}} $ \cite{Dutta:2010jh}.

\begin{figure}[ht]
\begin{center}
\includegraphics[width=6cm]{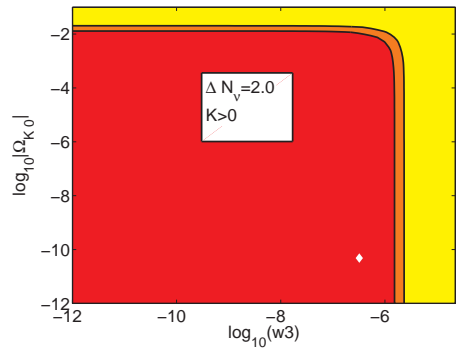}
\caption{ (Color Online) {\it{1$\sigma$ and  2$\sigma$ contour
plots of $\log_{10}(w_3)$ vs  $\log_{10}\vert\Omega_{ K0}\vert$
for $ K>0$ and $\dn=2.0$, using SNIa, BAO and CMB data.   The
white diamond marks the best-fit point. }}} \label{posk2}
\end{center}
\end{figure}

\begin{figure}[ht]
\epsfig{file=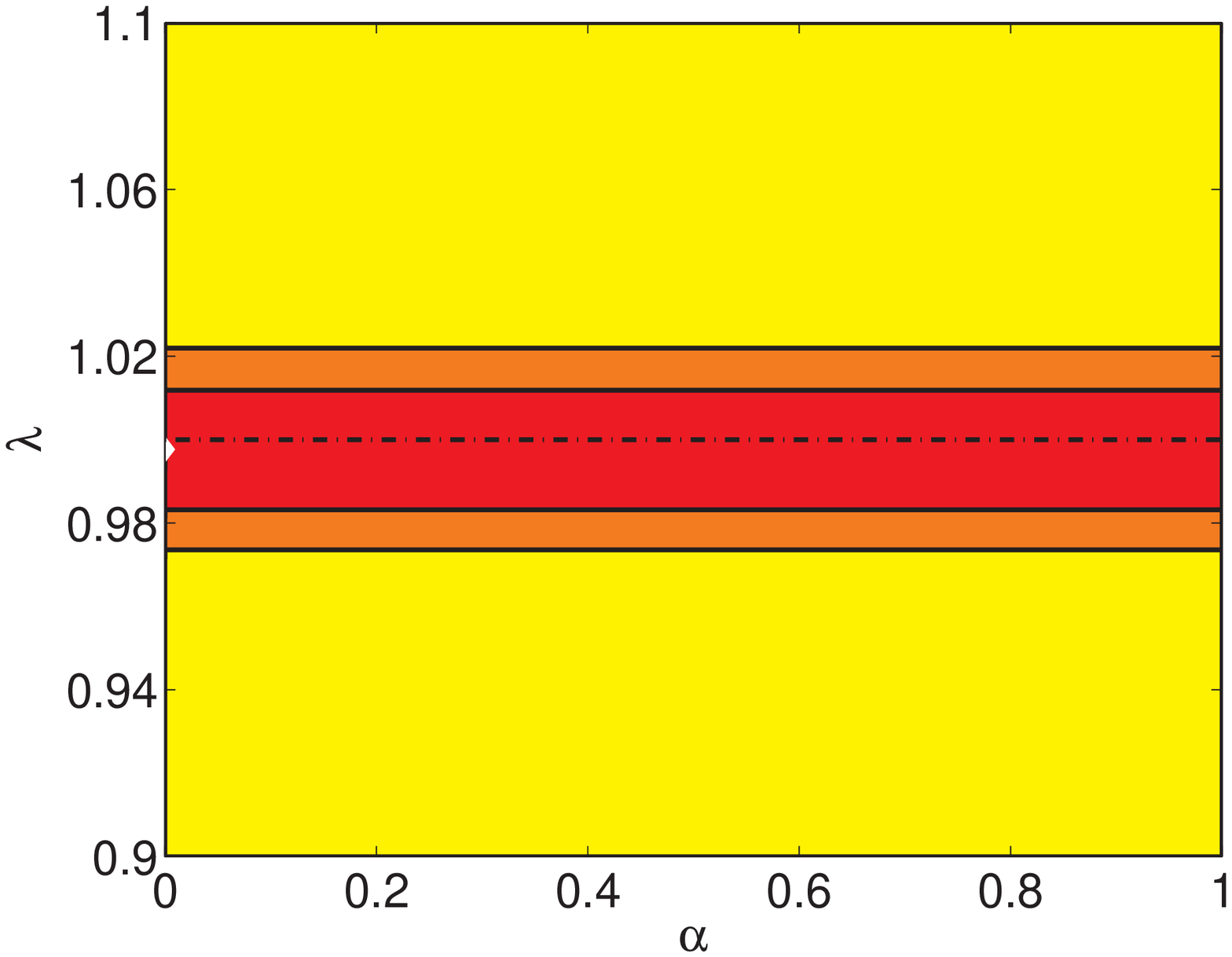,width=6 cm}\\ \epsfig{file=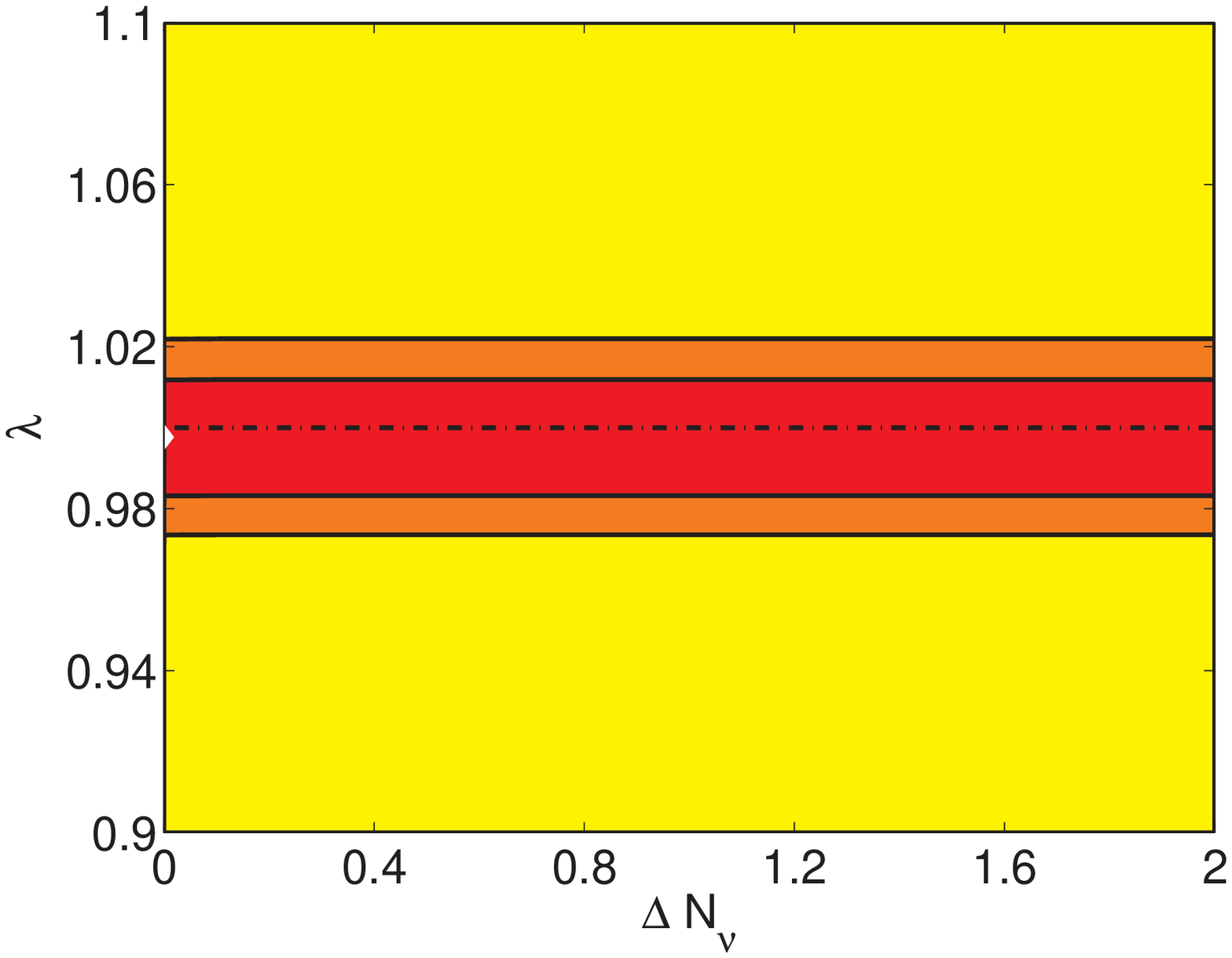,width=6 cm}\\
\epsfig{file=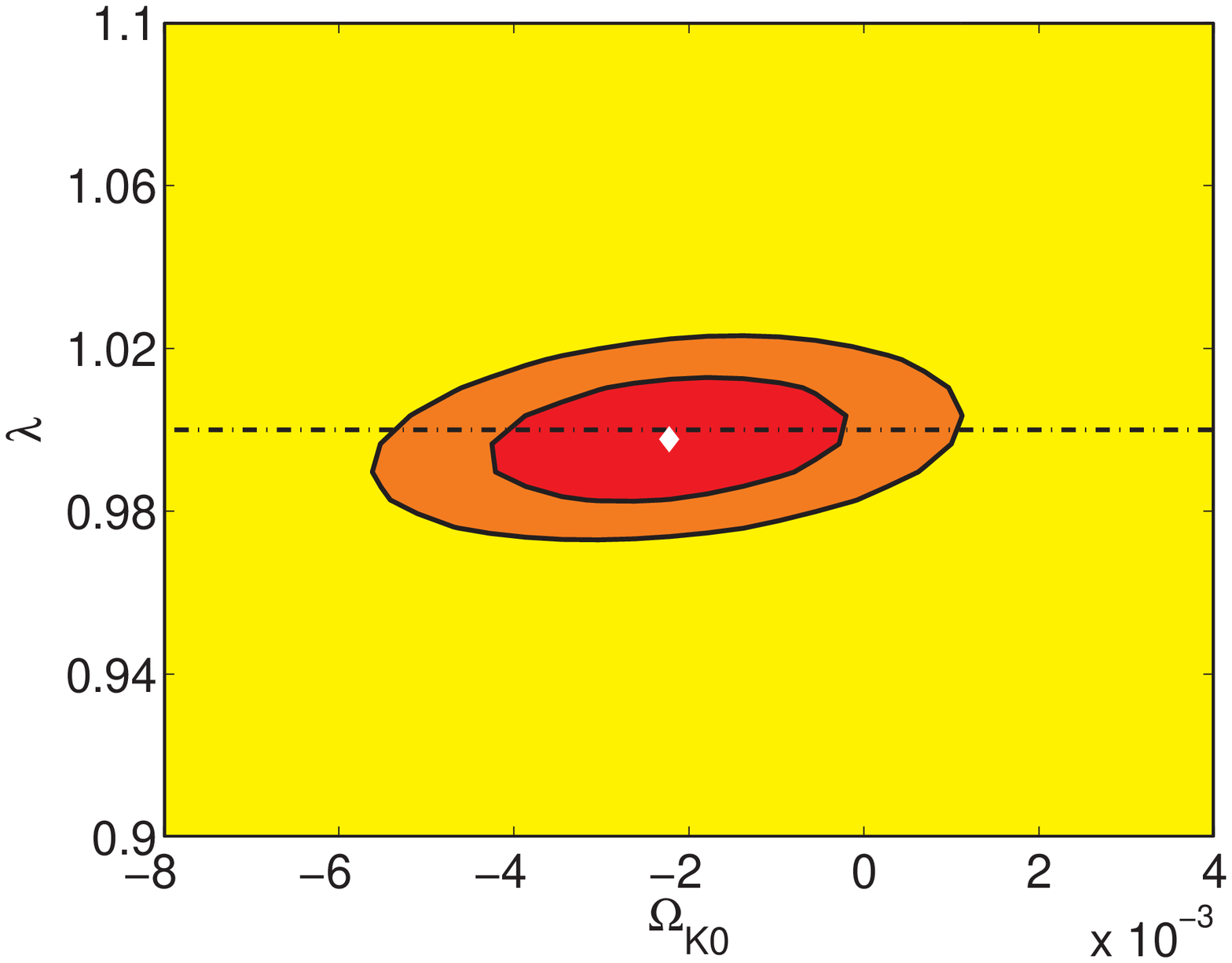,width=6 cm}
\caption{\label{bdbcontour}(Color online) {\it{Contour plots of
different pairs of free parameters in the beyond-detailed-balance
scenario,
        under SNIa, BAO and CMB observational data. In each case
         the parameters not included in the plots have been marginalized over. Color scheme as in Fig. \ref{dbcontour}.}}}
\end{figure}

We use relation (\ref{ndbcond2}) to eliminate $\w_4$ in favor of
$\alpha$ and $\dn$, and treat  $\lambda$,  $\alpha$, $\Omega_{K0}$
and $\dn$ as our free parameters, marginalizing over $H_0$,
$\Omega_{m0}$, $\Omega_{b0}$ and $\Omega_{r0}$. Using the combined
SNIa+CMB+BAO data, we construct likelihood contours for different
combinations of the above  parameters. Figure  \ref{posk2} depicts
the $1\sigma$ and $2\sigma$ $\w_3-\vert\Omega_{ K 0}\vert$
contours, for $\dn=2$, for positive  curvature, while Fig.
\ref{bdbcontour} depicts the $\lambda$-variation.
The approximate $1\sigma$ limits on the model parameters
$\sigma_i$ are presented in Table \ref{ndblimits0}.
\begin{table*}
    \centering
        \begin{tabular}{|c|c|c|c|c|c|c|}
        \hline
        $\sigma_0/(8\pi G)$&    \textbf{$\dn$} &  \textbf{$\Omega_{K0}$}        & \textbf{$\left(8\pi G/H_0^2\right)\sigma_1$} & $\sigma_2$       & \textbf{$\(8\pi G H_0^2\)\sigma_3$}          &  \textbf{$\sigma_4/(8\pi G)$}              \\\hline
       1/3& $0.1$          &     $(0,\, 0.01)   $          &   $(4.29,\,4.33)$ & $-6$     &         $(0,\, 0.03)$   &          $(-9.08\times 10^{-22},\,0)$                \\\hline
          1/3& $0.1$          &     $(-0.01,\, 0)   $          &   $(4.40,\,4.45)$ & $-6$      &         $(0,\, 0.81)$   &          $(0,\,5.66\times 10^{-22})$                \\\hline
         1/3& $2.0$          &     $(0,\, 0.04)   $          &   $(4.13,\,4.45)$ & $-6$      &         $(0,\, 0.01)$   &          $(-1.77\times 10^{-20},\,-2.62\times 10^{-21})$    \\\hline
          1/3&  $2.0$          &     $(-0.01,\, 0)   $          &   $(4.40,\,4.45)$  & $-6$    &         $(0,\, 0.23)$   &          $(-2.61\times 10^{-20},\,-1.16\times 10^{-20})$                \\\hline
        \end{tabular}
    \caption {1$\sigma$ limits on the parameter
     values for the beyond-detailed-balance scenario, for positive
     and negative curvature, and for two values of the effective neutrino
species parameter $\Delta N_\nu$ (see text).}
    \label{ndblimits0}
\end{table*}
Additionally, in Table \ref{ndblimits} we focus on the $1\sigma$
limits of $\alpha$ and $\lambda$.
\begin{center}
\begin{table}[ht]
    \centering
    \scalebox{.92}{
%           \scalebox{1.5}{
        \begin{tabular}{|c|c|c|c|}
        \hline
        \textbf{$\Omega_{K0}$} &    \textbf{$\dn$} &  \textbf{$\alpha$}       & \textbf{$\lambda$}  \\\hline
        $(-0.01,0.01)$         & $(0,2)$         & $(0,\, 1)$            &  $(0.98,\,1.01)$    \\\hline
        \end{tabular}}
    \caption {1$\sigma$ limits on the free parameters of the beyond-detailed-balance scenario. The cosmological parameters $\Omega_{m0}$, $\Omega_{b0}$, $\Omega_{r0}$ and $H_0$ have been marginalized over.}
    \label{ndblimits}
\end{table}
\end{center}
As we observe, in $1\sigma$ confidence the running parameter
$\lambda$ of Ho\v{r}ava-Lifshitz gravity is restricted to the
interval $|\lambda-1|\lesssim0.02$, for the entire allowed range
of $\omega_3$ (that is of $\sigma_3$). Finally, the best fit value
for $\lambda$ restricts $|\lambda-1|$ to much more smaller values,
namely $|\lambda_{b.f}-1|\approx0.002$.

\section{Phase-space analysis of Ho\v{r}ava-Lifshitz cosmology}
\label{leonmanos}

In this section we review the results of the phase-space and
stability analysis of Ho\v{r}ava-Lifshitz cosmology, with or
without the detailed-balance condition, following
\cite{Leon:2009rc}. We are interested in investigating the
possible late-time solutions, and in these solutions we calculate
various observable quantities, such are the dark-energy density
and equation-of-state parameters.

We start by transforming the cosmological equations into an
autonomous dynamical system \cite{Copeland:1997et}, introducing
suitable dimensionless variables which are combinations of the
model variables and parameters. Then we extract the critical
points of the autonomous system, and in order to determine their
stability we linearize it around them and we examine the
eigenvalues of the corresponding coefficient matrix of the
perturbation equations.

In the case where the detailed-balance condition is imposed, we
find that the universe can reach a bouncing-oscillatory state at
late times, in which dark-energy, behaving as a simple
cosmological constant, will be dominant. Such solutions   arise
purely from the novel terms of Ho\v{r}ava-Lifshitz cosmology, and
in particular the dark-radiation term proportional to $a^{-4}$ is
responsible for the bounce, while the cosmological constant term
is responsible for the turnaround.

In the case where the detailed-balance condition is abandoned, we
find that the universe reaches an eternally expanding solution at
late times, in which dark-energy, behaving like a cosmological
constant, dominates completely. Note that according to the initial
conditions, the universe on its way to this late-time attractor
can be an expanding one with non-negligible matter content,
independently of the specific form of the dark-matter content.
These features make this scenario a good candidate for the
description of our universe, in consistency with observations.
Finally, in this case the universe has also a probability to reach
an oscillatory solution at late times, if the initial conditions
lie in its basin of attraction.

\section{Bounce and Cyclic behavior}\label{bouncecyclic}

The possibility of late-time cyclic solutions that arose from the
phase-space analysis, makes us to investigate it in more detail.
Let us take a first look at how it is possible to obtain a
cosmological bounce in this framework \cite{Cai:2010zma}. In the
contracting phase we have $H<0$, while in the expanding one we
have $H>0$, and by making use of the continuity equations it
follows that at the bounce point $H=0$. Throughout this transition
$\dot H
> 0$.   On the other
hand, for the transition from expansion to contraction, that is
for the cosmological turnaround, we have $H>0$ before and $H<0$
after, while exactly on the turnaround point we have $H=0$.
Throughout this transition $\dot H < 0$.

The above conditions for a bounce and a turnaround can be easily
fulfilled in Ho\v{r}ava-Lifshitz cosmology, as we observe from the
two Friedmann equations (\ref{Fr1c}) and  (\ref{Fr2c}). In
particular, a cyclic scenario could be straightforwardly obtained
if we consider a negative dark radiation term and a negative
cosmological constant. During the expansion, the energy densities
of all components decrease, which is not the case for the
cosmological constant. Thus, its contribution will counterbalance
that of dark matter, triggering a turnaround, after which the
universe enters in the contracting phase. Then, after contraction
to sufficiently small scale factors the dark radiation term will
lead the universe to experience a bounce. Thus, the universe in
such a model indeed presents a cyclic behavior, with a bounce and
a turnaround at each cycle \cite{Cai:2009in}.

The absence of singularities in a cosmological scenario  is a
significant advantage. However, one must examine the proceeding of
fluctuations through the bounce. In general, non-relativistic
gravities, such is Ho\v{r}ava-Lifshitz one, are usually able to
recover Einstein's general relativity as an emergent theory at low
energy scales. Therefore, the cosmological fluctuations generated
in this model should be consistent with those obtained in standard
perturbation theory in the IR limit \cite{Brandenberger:2009yt}.
In particular, the perturbation spectrum presents a
scale-invariant profile, if the universe has undergone a
matter-dominated contracting phase \cite{Cai:2008qw}. However, the
non-relativistic corrections in the Ho\v{r}ava-Lifshitz action
could lead to a modification of the dispersion relations of
perturbations. This issue has been addressed in \cite{Cai:2009hc}
(see \cite{Mukohyama:2009gg} and references therein for the
perturbations of a pure expanding universe in Ho\v{r}ava-Lifshitz
cosmology), which shows that the spectrum in the UV regime may
have a red tilt in a bouncing universe. Moreover, the perturbation
modes cannot enter the UV regime in the scenario of matter-bounce.
Thus, the analysis of the cosmological perturbations in the IR
regime is quite reliable.

\section{A more realistic Ho\v{r}ava-Lifshitz dark energy}
\label{realde}

In section \ref{model} we formulated Ho\v{r}ava-Lifshitz
cosmology, in which one can define the effective dark energy
sector through (\ref{rhoDE}),(\ref{pDE}) in the detailed-balance
case, or through (\ref{rhoDEext}),(\ref{pDEext}) in the
beyond-detailed-balance case. Thus, one can straightforwardly
obtain the dark-energy equation-of-state parameter in both cases,
as $w_{DE}=p_{DE}/\rho_{DE}$. As can be immediately seen, in both
cases $w_{DE}$ lies above the phantom divide. However, according
to observations, $w_{DE}$ could have crossed $-1$ in the recent
cosmological past. Therefore, the question is wether we can
formulate an extension of Ho\v{r}ava-Lifshitz cosmology, in which
the dark energy equation-of-state parameter can experience the
phantom-divide crossing.

For this shake we allow for an additional scalar field, which will
contribute to the dark energy sector \cite{Saridakis:2009bv}
\footnote{ Note that one could alternatively generalize the
gravitational action of Ho\v{r}ava-Lifshitz gravity itself
\cite{Nojiri:2009th,Kluson:2009rk,Chaichian:2010yi}. }. Hence, we
add a second scalar $\sigma$, with action
\begin{eqnarray}
&&S_\sigma = \int dtd^3x \sqrt{g} N \left[
\frac{3\lambda-1}{4}\frac{\dot\sigma^2}{N^2}+\ \ \ \ \ \ \ \ \ \ \
\ \ \ \ \ \ \ \ \ \ \
 \ \ \ \   \ \right.\nonumber\\
&&\ \ \ \ \ \ \left.
h_1h_2\sigma\nabla^2\sigma-\frac{1}{2}h_2^2\sigma\nabla^4\sigma +
\frac{1}{2}h_3^2\sigma\nabla^6\sigma -V(\sigma) \right], \ \
\end{eqnarray}
where $V(\sigma)$ accounts for the potential term of the
$\sigma$-field and $h_i$ are constants. Assuming homogeneity, that
is $\sigma\equiv\sigma(t)$, its evolution equation will be given
by
\begin{eqnarray}\label{phidott}
\label{sdott}
 \ddot\sigma+3H\dot\sigma+\frac{2}{3\lambda-1}\frac{dV(\sigma)}{d\sigma}=0.
\end{eqnarray}
Additionally, it can be easily seen that its contribution to the
Friedmann equations of section \ref{model} will be the standard
scalar-field one, and thus one can absorb it in an extended dark
energy sector, with energy density and pressure given by:{{\small
\begin{eqnarray}\label{rhoDE22}
&&\rho_{DE}\equiv\frac{3\la-1}{4}\,\dot\sigma^2 +V(\sigma)
+\frac{3\kappa^2\mu^2K^2}{8(3\lambda-1)a^4}
+\frac{3\kappa^2\mu^2\Lambda ^2}{8(3\lambda-1)} \nonumber\\
\label{pDE22}
  &&p_{DE}\equiv\frac{3\la-1}{4}\,\dot\sigma^2
-V(\sigma) +\frac{\kappa^2\mu^2K^2}{8(3\lambda-1)a^4}
-\frac{3\kappa^2\mu^2\Lambda ^2}{8(3\lambda-1)}\nonumber
\end{eqnarray}}}
in the detailed-balance case, and by: {{\small
\begin{eqnarray}\label{rhoDEext22}
&&\rho_{DE}|_{_\text{non-db}}\equiv\frac{3\la-1}{4}\,\dot\sigma^2
+V(\sigma)+ \frac{\sigma_1}{6}+\frac{\sigma_3 K^2}{6a^4}
+\frac{\sigma_4 K}{6a^6}
\nonumber\\
&&\label{pDEext22}
p_{DE}|_{_\text{non-db}}\equiv\frac{3\la-1}{4}\,\dot\sigma^2
-V(\sigma) -\frac{\sigma_1}{6}+\frac{\sigma_3 K^2}{18a^4}
+\frac{\sigma_4 K}{6a^6}\nonumber
\end{eqnarray}}}
in the beyond-detailed-balance one. Note that the dark energy
density in both cases satisfies the usual conservation equation.

The aforementioned extended version of Ho\v{r}ava-Lifshitz dark
energy can have a very interesting phenomenology. Firstly, the
corresponding equation-of-state parameter $w_{DE}$ can be above
$-1$, below $-1$, or experience the $-1$-crossing during the
cosmological evolution, as can be straightforwardly seen by the
ratio $w_{DE}=p_{DE}/\rho_{DE}$. Thus, in this case, artifacts of
Ho\v{r}ava-Lifshitz gravity could be detected through dark energy
observations. However, one still cannot distinguish between this
model and alternative models that allow for the realization of
$w_{DE}<-1$ phase, such are modified gravity \cite{Nojiri:2003ft}
or models with phantom \cite{phant} or quintom fields
\cite{quintom}. However, note that in the present formulation the
additional scalar field is canonical, while in phantom and quintom
scenarios the scalar field is phantom, and thus with ambiguous
quantum behavior. The ability to describe the phantom phase and
the phantom crossing with a canonical scalar field is a
significant advantage of the scenario at hand, revealing the
capabilities of Ho\v{r}ava-Lifshitz cosmology.

\section{Perturbative instabilities in Ho\v{r}ava-Lifshitz gravity}
\label{instabil}

In the previous sections we showed the advantages of
Ho\v{r}ava-Lifshitz cosmology at the background level. However,
despite the capabilities of the scenario, our analysis does not
enlighten the discussion about the possible conceptual problems
and instabilities of Ho\v{r}ava-Lifshitz gravity,  nor it can
address the questions concerning the validity of its theoretical
background. Thus, in this section we are interested in performing
a detailed investigation of the gravitational perturbations of
Ho\v{r}ava-Lifshitz gravity, using it as a tool to examine its
consistency, studying both scalar and tensor sectors around a
Minkowski background \cite{Bogdanos:2009uj}.

We consider coordinate transformations of the form $x^{\mu} \to
\tilde x^{\mu}=x^{\mu}+\xi^{\mu}$. Under this transformation the
metric-perturbation around a given background changes as $\delta
\tilde g_{\mu \nu} = \delta g_{\mu \nu}-\nabla_{\mu}
\xi_{\nu}-\nabla_{\nu} \xi_{\mu}$.
 Therefore, the general
perturbations of the metric (\ref{metriciit})   read:
\begin{eqnarray}
\delta g_{00}&=&-2a^2\phi\nonumber\\
\delta g_{0i}&=&a^2\partial_i B+a^2Q_i\nonumber\\
\delta g_{ij}&=&a^2h_{ij}-a^2(\partial_i W_j+\partial_j W_i)-
2a^2 \psi \delta_{ij}+2 a^2
\partial_i
\partial_j E.\nonumber
 \end{eqnarray}
The vector modes are assumed to be transverse, that is $\partial_i
W^i =\partial_i Q^i=0$, while  the tensor mode is forced to be
transverse and traceless: $\partial_i h^{ij}=\delta^{ij}h_{ij}=0$.

Let us now discuss   the gauge fixing, which is required for the
action derivation and the determination of the physical degrees of
freedom. The projectability condition of Ho\v{r}ava gravity
\cite{hor3} requires that the perturbation of the lapse-function
$N$ depends only on time, thus $\phi\equiv\phi(t)$. This allows us
to ``gauge away'' the $\phi$- and $B$-perturbations, and also we
can   eliminate the $Q_i$ degree of freedom
\cite{Bogdanos:2009uj}. Therefore, the remaining degrees of
freedom are $\psi$, $E$, $W_i$ and $h_{ij}$. In summary, in the
aforementioned gauge we obtain
\begin{eqnarray}
\delta N &=& \delta N_i =0\nonumber\\
\delta_{ij} &=& h_{ij}-2\psi\delta_{ij}+2\partial_i\partial_j
E-(\partial_i W_j+\partial_j W_i).
 \end{eqnarray}
Note that since only perturbations imposed on the ``same-time''
spatial hypersurface are allowed, this is equivalent to a {\it
synchronous gauge} choice.

We now perturb the (prior to analytic continuation)
Ho\v{r}ava-Lifshitz gravitational action up to second order. After
non-trivial but straightforward calculations
\cite{Bogdanos:2009uj}, for the perturbed kinetic part of the
action  (\ref{acct}) we obtain
\begin{eqnarray}
\delta S^{(2)}_K=\int dt d^3x \frac{2}{\kappa^2}\Big[\frac{1}{4}
\dot h_{ij} \dot h^{ij} &+&(1-3 \lambda) \left( 3 \dot \psi^2 - 2
\dot \psi \nabla^2 \dot E \right)\nonumber\\
&+& (1-\lambda) \dot E \nabla^4 \dot E \Big],
 \label{DSK}
  \end{eqnarray}
  while for the perturbed potential part
  we acquire
\begin{eqnarray} \delta S_V^{(2)}  &=& \int d
td^3 x\left[ {\frac{{\kappa ^2 }} {{8w^4 }}h_{ij} \nabla ^6 h^{ij}
+ \frac{{\kappa ^2 \mu }} {{8w^2 }}\epsilon^{ijk} h_{il} \partial
_j \nabla ^4 h_k^l   } \right.\: \nonumber \\
&& \;\;\;\;\;\;\;\;\;\;\;\;\;\;\;- \frac{{\kappa ^2 \mu ^2 }}
{{32}}h_{ij} \nabla ^4 h^{ij}+ \frac{{\kappa ^2 \mu ^2 \Lambda }}
{{32(1 - 3\lambda )}}h_{ij} \nabla ^2 h^{ij}  \: \nonumber \\
&& \;\;\;\;\;\;\;\;\;\;\;\;\;\;\; - \frac{{\kappa ^2 \mu ^2 (1 -
\lambda )}} {{4(1 - 3\lambda )}}\psi \nabla ^4 \psi -
\frac{{\kappa ^2 \mu ^2 \Lambda }}
{{4(1 - 3\lambda )}}\psi \nabla ^2 \psi   \nonumber \\
&& \;\;\;\;\;\;\;\;\;\;\;\;\;\;\; + \frac{{27\kappa ^2 \mu ^2
\Lambda ^2 }} {{16(1 - 3\lambda )}}\psi ^2  - \frac{{9\kappa ^2
\mu ^2 \Lambda ^2 }} {{8(1 - 3\lambda )}}\psi \nabla ^2 E\nonumber \\
&& \;\;\;\;\;\;\;\;\;\;\;\;\;\;\;  \left. { + \frac{{3\kappa ^2
\mu ^2 \Lambda ^2 }} {{16(1 - 3\lambda )}}E\nabla ^4 E} \right]
\label{DSV}.
 \end{eqnarray}

\subsection{Scalar perturbations}

As can be observed from (\ref{DSK}),(\ref{DSV}) the action for
scalar perturbations includes the two modes  $E$ and $\psi$, and
  their equations of motion read:{\small{
 \begin{equation}
  \frac{8} {{\kappa ^2
}}\ddot E + \frac{{\kappa ^2 \mu ^2 \left( {1 - \lambda }
\right)}} {{2\left( {1 - 3\lambda } \right)}}\nabla ^2 \psi  +
\frac{{\kappa ^2 \mu ^2 \Lambda }} {{2\left( {1 - 3\lambda }
\right)}}\psi  = 0\,
 \label{alphaeq}
 \end{equation}
\begin{eqnarray}
 \frac{8} {{\kappa ^2 }}\frac{{1 - 3\lambda }} {{1
- \lambda }}\ddot \psi &-& \frac{{9\kappa ^2 \mu ^2 \Lambda ^2 }}
{{4\left( {1 - \lambda } \right)\left( {1 - 3\lambda }
\right)}}\psi  + \frac{{3\kappa ^2 \mu ^2 \Lambda ^2 }} {{4\left(
{1 - \lambda } \right)\left( {1 - 3\lambda } \right)}}\nabla ^2 E
\nonumber \\  &+& \frac{{\kappa ^2 \mu ^2 \left( {1 - \lambda }
\right)}} {{2\left( {1 - 3\lambda } \right)}}\nabla ^4 \psi+
\frac{{\kappa ^2 \mu ^2 \Lambda }} {{2\left( {1 - 3\lambda }
\right)}}\nabla ^2 \psi  = 0.
 \label{betaeq}
 \end{eqnarray}}}
As can be seen these two equations are coupled, not allowing for a
straightforward stability investigation. However, we can still
acquire information about the stability of the configuration by
studying it at high and low momenta. Taking the IR limit of
  (\ref{betaeq}), that is considering the low-$k$
behavior, it reduces to
 \begin{equation}
  \frac{8} {{\kappa ^2
}}\frac{{1 - 3\lambda }} {{1 - \lambda }}\ddot \psi  -
\frac{{9\kappa ^2 \mu ^2 \Lambda ^2 }} {{4\left( {1 - \lambda }
\right)\left( {1 - 3\lambda } \right)}}\psi  = 0 \label{psi1}.
\end{equation}
 Thus, this decoupled equation  acts as a low-momentum
equation of motion for the scalar field $\psi$. A straightforward
observation from (\ref{psi1}) is that it leads to a ghost-like
behavior, since it leads to
  the   dispersion relation
 \begin{equation}
\omega^2\equiv m^2 = - \frac{{9\kappa ^4 \mu ^2 \Lambda ^2 }}
{{32\left( {1 - 3\lambda } \right)^2 }} < 0 \label{masseq},
 \end{equation}
which induces instabilities, regardless of the $\lambda$-value
 and of the sign of the cosmological constant.
Now, for high $k$,   (\ref{betaeq}) reduces to
  \begin{equation}
\frac{8} {{\kappa ^2 }}\frac{{1 - 3\lambda }} {{1 - \lambda
}}\ddot \psi  + \frac{{\kappa ^2 \mu ^2 \left( {1 - \lambda }
\right)}} {{2\left( {1 - 3\lambda } \right)}}\nabla ^4 \psi  = 0.
\label{psi2}
 \end{equation}
Therefore,    (\ref{psi2}) yields the high-$k$ dispersion
relation:
 \begin{equation}
   \omega ^2  \equiv
  \frac{{\kappa ^4 \mu ^2 }}
{{16}}\left( {\frac{{1 - \lambda }} {{1 - 3\lambda }}} \right)^2
k^4.
 \label{omegascalar}
  \end{equation}

\subsection{Tensor perturbations}

Let us now examine the tensor perturbations. Their action can be
extracted from (\ref{DSK}),(\ref{DSV}) and therefore the graviton
equation of motion writes as
\begin{eqnarray}
\ddot h^{ij}- \frac{{\kappa ^4 \mu ^2 \Lambda }} {{16(1 - 3\lambda
)}}\nabla ^2 h^{ij}  - \frac{\kappa^4}{4 w^4} \nabla^6 h^{ij} -
\frac{\kappa^4 \mu}{4 w^2} \epsilon^{ilk} \partial_l \nabla^4
h^j_k \nonumber\\
+ \frac{\kappa^4 \mu^2}{16} \nabla^4 h^{ij} = 0.\ \ \
\end{eqnarray}
 Assuming graviton propagation along the $x^3$
direction, that is $k_i=k^i=(0,0,k)$, the $h_{ij}$ can be written
as usual in terms of the Left and Right polarization components,
and thus we derive the two equations for the different
polarizations
\begin{eqnarray}
-\omega^2\tilde{h}_{L,R} + c^2k^2\tilde{h}_{L,R} +
\frac{\kappa^4\mu^2}{16}k^4\tilde{h}_{L,R} \pm
\frac{\kappa^4\mu}{4w^2}k^5\tilde{h}_{L,R} \nonumber\\+
\frac{\kappa^4}{4w^4}k^6\tilde{h}_{L,R} = 0,
\end{eqnarray}
where the plus and minus branches correspond to Left-handed and
Right-handed modes respectively. In this relation we have
identified the light speed from the low $k$ regime as $ c^2  =
 {\kappa ^4 \mu ^2 \Lambda }/[{16(1 - 3\lambda )}].
$ The above equation system   accepts a non-trivial solution only
if the corresponding determinant is zero, which leads to the
dispersion relation
\begin{equation}
\label{omegatensor}
 \omega ^2  = c^2 k^2  + \frac{{\kappa ^4 \mu ^2
}} {{16}}k^4  \pm \frac{{\kappa ^4 \mu }} {{4w^2 }}k^5  +
\frac{{\kappa ^4 }} {{4w^4 }}k^6.
\end{equation}

\subsection{Beyond Detailed Balance}
\label{nodetbal}

  In order to avoid possible accidental artifacts of the
detailed-balance condition, in this subsection we extend the
investigation beyond detailed balance. As a demonstration, and
without loss of generality, we consider a
detailed-balance-breaking term of the form $\nabla _i R_{jk}
\nabla ^i R^{jk} $. Thus, the corresponding contribution to the
action will be \cite{Bogdanos:2009uj}
\begin{equation}
 \delta S^{\left( 2 \right)} _{bdb}  = \eta \int
{dtd^3 x\left( { - \frac{1} {4}h_{ij} \nabla ^6 h^{ij}  - 6\psi
\nabla ^6 \psi } \right)},
\end{equation}
where $\eta$ is an additional parameter. It is straightforward to
calculate the modifications that $S^{\left( 2 \right)} _{bdb}$
brings to the dispersion relations for scalar and tensor
perturbations obtained above (expressions (\ref{omegascalar}) and
(\ref{omegatensor}) respectively).
 The extended
dispersion relations read:
\begin{equation}
\label{scalarnodb} \omega ^2  \sim \frac{{\kappa ^2 \left( {1 -
\lambda } \right)^2 }} {{16\left( {1 - 3\lambda } \right)^2
}}\,k^4 - \frac{{3\kappa ^2 \left( {1 - \lambda } \right)}}
{{2\left( {1 - 3\lambda } \right)}}\eta k^6
\end{equation}
 for scalar perturbations (UV-behavior), and
\begin{equation}
\label{tensornodb} \omega ^2  = c^2 k^2  + \frac{{\kappa ^4 \mu ^2
}} {{16}}k^4  \pm \frac{{\kappa ^4 \mu }} {{4w^2 }}k^5  + \left(
\frac{{\kappa ^4 }} {{4w^4 }} - \frac{\kappa^{2}\eta}{2}\right)
k^6
\end{equation}
for tensor perturbations. As was expected, the
detailed-balanced-breaking term  modifies mainly the UV regime of
the theory.

\subsection{Instabilities} \label{discussion}

Concerning  the scalar perturbations, as was mentioned above
(\ref{psi1}),(\ref{psi2}) leads to instabilities. This unstable
behavior cannot be cured by simple tricks such as analytic
continuation  of the form $ \mu\rightarrow i\mu~,~~w^2\rightarrow
-iw^2 4 $ \cite{Lu:2009em}, since in that case we
straightforwardly see that the UV behavior is spoiled (see
(\ref{omegascalar})) and thus instabilities re-emerge at high
energies. Even in this case though, we cannot evade the
instability coming from the negative mass term, and thus IR
instabilities persist  as long as we have a non-vanishing
cosmological constant. Finally, concerning the tensor sector, from
(\ref{omegatensor}) we see that if we desire a well-behaved UV
regime we cannot impose the analytic continuation.

Proceeding to the relaxation of the detailed-balance condition, a
crucial observation is that the ghost instability of the scalar
mode arises from the kinetic term of the action and thus the
breaking of detailed balance, which affects the potential term,
will not alter the aforementioned scalar-instabilities results.

\section{Healthy extensions of Ho\v{r}ava-Lifshitz gravity}
\label{healthy}

In the previous section we saw that Ho\v{r}ava-Lifshitz gravity in
its simple version, with or without the detailed-balance version,
suffers from instabilities and pathologies that cannot be cured.
It is thus necessary to try to construct suitable extensions that
are free of such problems.

A quite general power-counting renormalizable action is
\cite{Kiritsis:2009vz}:
 \be S=S_{kin}+S_{1}+S_{2}+S_{new}
\label{q10},\ee with
 \be S_{kin}=\alpha
\int dtd^3x\sqrt{g}N\!\!\left[(K_{ij}K^{ij}\!-\!\l K^2)\right]
\label{q4}\nonumber\ee \be S_{1}=\int
dtd^3x\sqrt{g}N\left[\gamma_0 {\e^{ijk}\over
\sqrt{g}}R_{il}\na_j{R^l}_{k} \!+\!\zeta R_{ij}R^{ij}\!+\!\eta
R^2\!+\!\xi R\!+\!\sigma\!\right]
 \label{q5}\nonumber
 \ee
  \begin{eqnarray} S_2=\int dtd^3x\sqrt{g}N\left[ \beta_0
C_{ij}C^{ij}+\beta_1 R\square
R+\beta_2R^3\right.\nonumber\\
\left.+\beta_3RR_{ij}R^{ij}+\beta_4 R_{ij}R^{ik}{R^{j}}_k\right]
\label{q6}\nonumber
\end{eqnarray}
  \begin{eqnarray} S_{new}=\int
dtd^3x\sqrt{g}N\left[a_1 (a_ia^i)+a_2 (a_ia^i)^2+a_3R^{ij}a_ia_j
\right.\nonumber\\
\left.+a_4R\nabla_i a^i+a_5\nabla_ia_j\nabla^ia^j+ a_6\nabla^i a_i
(a_ja^j)+\cdots \right].\ \ \label{q7}\end{eqnarray} Thus, apart
from the known kinetic, detailed-balance and
beyond-detailed-balance combinations that constitute the
Ho\v{r}ava-Lifshitz gravitational action, in (\ref{q7}) we have
added a new combination, based on the term \cite{Blas:2009qj}:
 \be
 a_i\equiv {\partial_iN\over N},
 \label{q11}
 \ee
 which breaks the projectability condition,
and the ellipsis in (\ref{q7}) refers to dimension six terms
involving $a_i$ as well as curvatures.

Such a new combination of terms seems to alleviate the problems of
Ho\v{r}ava-Lifshitz gravity, although there could still be some
ambiguities \cite{Papazoglou:2009fj}. Therefore, one should repeat
all the investigations of the present work, for this extended
version of the theory.

\section{Conclusions}
\label{conclusions}

In this work we reviewed some general aspects of
Ho\v{r}ava-Lifshitz cosmology. Formulating the basic version of
Ho\v{r}ava-Lifshitz gravity, with or without the detailed-balance
condition, we extracted the cosmological equations. We used
observational data in order to constrain the parameters of the
theory, and amongst others we saw that the running parameter
$\lambda$, that determines the flow between the IR and the UV, is
indeed restricted in a very narrow window around its IR value 1.
Through a phase-space analysis we extracted the late-time stable
solutions, which are independent of the initial conditions, and we
saw that eternal expansion, or bouncing and cyclic behavior, can
arise naturally. Formulating the effective dark energy sector we
showed that Ho\v{r}ava-Lifshitz cosmology can describe the phantom
phase, without the use of a phantom field. However, performing a
detailed perturbation analysis, we showed that Ho\v{r}ava-Lifshitz
gravity in its basic version, suffers from instabilities. Thus,
one should try to construct suitable generalizations, that are
free from pathologies, and then repeat all the above steps of
cosmological analysis. Such a task proves to be hard, but it is
necessary if we desire Ho\v{r}ava-Lifshitz gravity to be a
candidate for the description of nature.

\begin{acknowledgments}
The author wishes to thank the Physics Department of National
Tsing Hua University of Taiwan, for the hospitality during the
preparation of this work.
\end{acknowledgments}

\end{document}